\documentclass[12pt]{article}

\usepackage{amsmath}
\usepackage[dvips]{graphics}
\usepackage{latexsym}

\DeclareFontFamily{OT1}{rsfs}{}
\DeclareFontShape{OT1}{rsfs}{m}{n}{ <-7> rsfs5 <7-10> rsfs7 <10->
rsfs10}{} \DeclareMathAlphabet{\mycal}{OT1}{rsfs}{m}{n}
\def\scri{{\mycal I}}%

\begin{document}
\newcommand{\bea}{\begin{eqnarray*}}
\newcommand{\eea}{\end{eqnarray*}}
\newcommand{\bean}{\begin{eqnarray}}
\newcommand{\eean}{\end{eqnarray}}
\newcommand{\eq}[1]{(\ref{#1})} 

\newcommand{\tri}{\delta}
\newcommand{\grad}{\nabla}
\newcommand{\pa}{\partial}
\newcommand{\mf}{\mathbf}

\newcommand{\om}{\omega}
\newcommand{\omo}{\omega_0}
\newcommand{\ep}{\epsilon}
\newcommand{\nonu}{\nonumber}
\newcommand{\scrip}{\scri^{+}}
\newcommand{\hp}{{\cal H^+}}

\newcommand{\lxi}{{\cal L}_\xi}
\newcommand{\lt}{{\cal L}_t}
\newcommand{\lchi}{{\cal L}_\chi}
\newcommand{\psig}{{\partial\Sigma}}
\newcommand{\dbar}{\bar\delta}
\newcommand{\La}{\Lambda}  
\newcommand{\sh}{S_{\cal H}}

\title{First law of black hole mechanics in Einstein-Maxwell and
  Einstein-Yang-Mills theories }     
\author{Sijie Gao \\
Centro Multidisciplinar de Astrof\'{\i}sica - CENTRA,\\
Departamento de F\'{\i}sica, Instituto Superior T\'ecnico,\\
Av. Rovisco Pais 1, 1049-001 Lisboa, Portugal}

\maketitle

\begin{abstract}
The first law of black hole mechanics is derived from the
Einstein-Maxwell Lagrangian by comparing two infinitesimally nearby
stationary black holes. With similar arguments, the first law of black
hole mechanics in Einstein-Yang-Mills  theory is also derived.

\end{abstract}
\maketitle

\section{Introduction} 
According to the ``no hair'' theorem, a general stationary black hole is
a charged and rotating black hole.  The first law of black hole
mechanics shows that the first order variations of the area $A$, mass
$M$, angular momentum $J$, and charge $Q$ are related by   
\bean
\frac{1}{8\pi}\kappa \delta A =\delta M- \Omega_{H}\delta
J-\Phi_{bh}\delta Q, \label{first}
\eean
where $\kappa$ denotes the surface gravity of the black hole,
$\Omega_H$ denotes the angular velocity of the horizon, and 
$\Phi_{bh}$ denotes the electrostatic potential of the horizon.
There are two versions of this law referred to by Wald \cite{thinbook}
as the ``physical process version'' and the ``equilibrium state
version.'' The ``physical process version'' of the first law is
obtained by changing a stationary black hole by some (infinitesimal)
physical process. The black hole is assumed to have settled down to
a new stationary final state. Then Eq. \eq{first} is derived by
comparing the final state of the black hole with the initial one
\cite{mine}. The ``equilibrium state'' version of the first law simply
compares the areas of two infinitesimally nearby stationary black hole
solutions. The original derivation was given by Bardeen {\em et al.}
\cite{bardeen}. However, since only a perfect fluid in circular
orbit around a black hole was considered, the first law in
\cite{bardeen} has a different form from Eq.\eq{first}. A simple derivation
in a general manner was given by Iyer and Wald \cite{iyer1} from the
Lagrangian formulation of general relativity. The derivation makes
essential use of the bifurcation two-sphere where the horizon Killing
vector field vanishes. This treatment requires that all fields be
smooth on the bifurcation surface, and consequently the
``potential-charge'' term does not appear explicitly in the first law.
The first 
task of this paper is to extend the work of \cite{iyer1} to a general
charged and rotating black hole where fields are not necessarily smooth through
the horizon.  The major
modification is that, instead of choosing the bifurcation surface as
the boundary of a hypersurface extending to spatial infinity, we
replace it with any cross section of the event horizon to the future
of the bifurcation surface (if one exists). We require that only the
pullback \cite{gr} of the vector potential $A_a$ to the horizon in the
future of the bifurcation surface be smooth. 
Now we present such an example. The
vector potential in the Reissner-Nordstr\"om spacetime is given by \cite{gr}
\bean
A_a=-\frac{Q}{r}(dt)_a \label{vp}.
\eean
To see the behavior of $A_a$ on the horizon, we introduce the
Kruskal coordinates $(U,V)$:
\bean
U&=&-e^{-\kappa u} \label{cu},\\
V&=&e^{\kappa v} \label{cv},
\eean
where 
\bean
u&=& t-r_* \label{u},\\
v&=&t+r_* \label{v},
\eean
In terms of $(U,V)$, $A_a$ can be written as
\bean 
A_a=-\frac{Q}{2\kappa r}\left[-\frac{1}{U}(dU)_a+\frac{1}{V}(dV)_a\right].
\label{auv}
\eean
We see immediately that $A_a$ is divergent at the bifurcation
$U=V=0$. Although $A_a$ is divergent on the future horizon $U=0$, $V>0$,
the pullback of $A_a$ to the future horizon (the restriction of $A_a$
to vectors tangent to the horizon)  is smooth. 
Since $A_a$ falls off as $1/r$ at infinity, it will have no
contribution to the canonical energy ${\cal E}$. 
As we shall see, the
charge term in Eq.\eq{first} emerges as an integration on  the
horizon. This modification also enables us to apply the result to  
black holes without a bifurcation surface, such as  extremal black
holes. A vector potential which is smooth through the horizon can 
easily be constructed by the gauge transformation 
\bean
\tilde A_a = -\frac{Q}{r}(dt)_a +\frac{Q}{ r_+}(dt)_a \label{sg}
\eean
where $r_+$ is the radial coordinate of the event horizon. Since $A_a$
is smooth through the horizon (identically zero), the
potential-charge term will not appear in the integral over the
horizon. However, $\tilde A_a$ in Eq.\eq{sg} does not drop to zero at
infinity; the potential-charge term will arise from infinity as
part of the canonical energy.

The second task of this paper is to generalize the method above to 
Einstein-Yang-Mills (EYM) black holes. The discovery of ``colored  black
Holes,'' such as black hole solutions in the   Einstein-Yang-Mills
theory, has been a great challenge to the traditional ``no hair''
conjecture. The first law of black-hole mechanics in the EYM case was
discussed by Sudarsky and Wald \cite{sw} and the following result 
was obtained:
\bean
\frac{1}{8\pi}\kappa \delta A =\delta M+V\delta Q^\infty - \Omega_{H}\delta
J,\label{farym}
\eean
where $V$ and $Q^\infty$ are the Yang-Mills potential and the charge
evaluated at infinity. The presence of this term is due to the
non-Abelian nature of the Yang-Mills field. The calculation also makes
use of the bifurcation two-sphere and all fields are
required to be smooth there. Again, we make no reference to the
bifurcation surface, and an additional surface term evaluated on any
cross section of the horizon is found [see \eq{firym}].

\section{First order variation of stationary spacetimes}
In this section, we briefly introduce a general variation theory
for stationary spacetimes in the framework of \cite{iyer1}.
We start with the general issue of calculating the first order
variation of conserved quantities. Consider a
diffeomorphism covariant theory in four dimensions derived from a
Lagrangian ${\mf L}$, where the dynamical fields consist of a Lorentz
signature metric $g_{ab}$ and other fields $\psi$. We follow the
notational conventions of \cite{iyer1}, and, in particular, we 
collectively refer to $(g_{ab},\psi )$ as $\phi$ and use boldface
letters to denote differential forms. According to \cite{iyer1}, the
first order variation of the Lagrangian can always be expressed as
\bean
\delta {\mf L} ={\mf E}(\phi)\delta\phi +d {\mf\Theta}(\phi, \delta\phi)
\label{dlf}
\eean
where ${\mf E }(\phi)$ is locally constructed out of $\phi$ and its
derivatives and ${\mf \Theta}$ is locally constructed out of $\phi$,$\delta
\phi$ and their derivatives. The equations of motion can then be read
off as
\bean
{\mf E}(\phi)=0. \label{eqom}
\eean
The symplectic current three-form ${\mf \omega}$ is defined by
\bean
\mbox{\boldmath $\omega$} (\phi, \delta_1\phi, \delta_2\phi)=\delta_1\mf
{\Theta} (\phi, 
\delta_2\phi)-\delta_2\mf{\Theta} (\phi, \delta_1\phi).
\label{sym}
\eean
The Noether current three-form associated with a smooth vector field $\xi$
is defined by 
\bean
{ \cal  J}=\mf\Theta (\phi,\lxi \phi)-\xi\cdot {\mf L}, \label{defj}
\eean
where ``$\cdot$'' denotes contraction of the vector field $\xi$ into
the first index of ${\mf L}$. A simple calculation yields 
\bean
d {\cal J}=-\mf E_\phi\lxi\phi.    \label{dje}
\eean
It was proved in the Appendix of \cite{iyer2} that there exists a
Noether charge two-form ${\mf Q}$, which is locally constructed from
$\phi$,$\xi^a$ and their derivatives, such that
\bean
{\cal J[\xi]}=d{\mf Q}[\xi]+\xi^a{\mf C}_a \label{jdqc}
\eean 
where $\mf C_a$ is a three-form and $\mf C_a=0$ when the equations of
motion are satisfied. Now suppose that the spacetime satisfies
asymptotic conditions at infinity corresponding to ``case I'' of \cite{waldz}
and that $\xi^a$ is an asymptotic symmetry. Then there exists a
conserved quantity $H_\xi$, associated with $\xi^a$. Let $\delta \phi$
satisfy the linearized equations of motion in the neighborhood of
infinity. Then $\delta H_\xi$ is given by \cite{waldz}
\bean
\delta H_\xi=\int_\infty (\dbar\mf Q[\xi]-\xi\cdot\mf\Theta).
\label{dhg}
\eean
Since $\xi^a$ is treated as a fixed background, it should not be varied
in the expression above.  So we used  ``$\dbar$'' to denote the
variation that has no effect on $\xi^a$, in distinction to the total
variation  ``$\delta$.'' 
Let $\Sigma$ be a hypersurface  that extends to infinity and has an
inner boundary $\partial \Sigma$. Now we consider the case where
$\xi^a$ is a symmetry of all the dynamical fields, i.e., $\lxi
\phi=0$, and $\delta \phi$ satisfies the linearized equations of
motion. Then Eq.(76) in \cite{iyer1} shows that the integral in
\eq{dhg} over infinity can be turned into one on the inner boundary, i.e.,
\bean
\delta H_\xi=\int_{\partial\Sigma} (\dbar\mf Q[\xi]-\xi\cdot\mf\Theta).
\label{dhob}
\eean
When $\xi^a$ is taken to be an asymptotic time translation $t^a$
and rotation $\phi^a$, respectively, we obtain the variations of
canonical energy ${\cal E}$ and canonical angular momentum $J$ \cite{iyer1}:
\bean
\delta{\cal E}&=&\int_\infty (\dbar\mf
Q[t]-t\cdot\mf\Theta)\label{ene}, \\
\delta J &=&-\int_\infty (\dbar\mf
Q[\varphi]-\varphi\cdot\mf\Theta)\label{angu}. 
\eean

\section{The first law of black hole mechanics in EM theory} 
We now specialize to Einstein-Maxwell theory. The dynamical fields are
$(g_{ab}, A_a)$ and the Einstein-Maxwell Lagrangian is 
\bean
{\mf L} = \frac{1}{16\pi}(\mbox{\boldmath$ \ep$} R-\mbox{\boldmath$
 \ep$}g^{ac}g^{bd}F_{ab}F_{cd}). \label{eml}
\eean
The Noether charge two-form $\mf Q$ and $\mf \Theta$
have been calculated in \cite{mine} as 
\bean
Q_{ab}=-\frac{1}{16\pi}\ep_{abcd}\grad^c\xi^d-\frac{1}{8\pi}
\ep_{abcd} F^{cd}A_e\xi^e \label{qan}
\eean
and 
\bean
\Theta_{abc}(\phi, \delta\phi)=\frac{1}{16\pi}\ep_{dabc}v^d, \label{theab}
\eean
where 
\bean
v_d=\grad^e\delta g_{de}-g^{fe}\grad_d \delta g_{fe}-4F_d^{\ b} \delta
A_b. 
\eean
Let $(g_{ab}, A_a)$ be a stationary solution to the Einstein-Maxwell
equations derived 
from the Lagrangian \eq{eml}. If the black hole possesses a bifurcation
surface, we require 
that the pullback of  $A_a$ to the future of
the bifurcation surface be smooth, but not necessarily smooth on the
bifurcation surface.  Let 
\bean
\xi^a=t^a+\Omega_H\varphi^a \label{hork}
\eean
denote the horizon Killing field of this black
hole \cite{thinbook}. Let $\Sigma$ be an asymptotic hypersurface
which terminates on the portion of the event horizon ${\cal H}$ to the
future of the bifurcation surface. Denote the cross section on  the
horizon by $\sh$, 
which is the inner boundary of $\Sigma$. Now consider a stationary
perturbation $\delta \phi$ that generates a slightly different
stationary axisymmetric black hole. When comparing two spacetimes,
there is a certain freedom in which points are chosen to
correspond. We shall adopt the gauge choice in \cite{bardeen}, i.e.,
we make the hypersurface $\Sigma$, the event horizons, and the Killing vectors
$t^a$ and $\varphi^a$ the same in the two solutions. Thus,
\bean
\delta t^a&=&\delta \varphi^a =0, \\
\delta \xi^a&=&\delta\Omega_H \varphi^a. \label{gauge}
\eean
Although the conditions above cannot be imposed on the bifurcation
surface where $\xi^a$ vanishes, our derivation will not be affected
since we shall make no use of the bifurcation surface.
If we assume that both $A_a$ and $\delta A_a$  fall off as fast as
 $1/r$ at infinity, as in the case in the introduction, then the
EM field contributes  
to neither $\delta {\cal E}$ nor $\delta J$ in Eqs.\eq{ene} and
\eq{angu}. Thus the variation of the canonical energy is the same as
that of the Arnowitt-Deser-Misner (ADM) mass $M$ and we shall rewrite  $\delta {\cal E}$ as
$\delta M$.
Combining Eqs. \eq{ene}, \eq{angu}, \eq{hork} and \eq{dhob}, we have
\bean
\delta M -\Omega_H\delta J =\int_{\sh}(\dbar \mf
Q[\xi]-\xi\cdot\mf \Theta \label{lefr})
\eean
Now we concentrate on the right-hand-side of Eq. \eq{lefr}. We shall
consider the contributions from the gravitational field and the EM
field separately. From Eq. \eq{qan}, we split $\mf Q$ as 
\bean
Q_{ab}=Q_{ab}^{GR}+Q_{ab}^{EM},
\eean
where
\bean
Q_{ab}^{GR}&=&-\frac{1}{16\pi}\ep_{abcd}\grad^c\xi^d  \label{qgr},\\
Q_{ab}^{EM}&=&-\frac{1}{8\pi} \ep_{abcd} F^{cd}A_e\xi^e.  \label{qem}
\eean
Similarly, we rewrite $\mf \Theta$ as 
\bean
\Theta_{abc}=\Theta_{abc}^{GR}+\Theta_{abc}^{EM},
\eean
where
\bean
\Theta_{abc}^{GR}&=&\frac{1}{16\pi}\ep_{dabc}g^{dh}(\grad^e\delta, 
g_{he}-g^{fe}\grad_h \delta g_{fe} ).\\
\Theta_{abc}^{EM} &=&\frac{1}{16\pi}\ep_{dabc}(-4F^{d b})\delta A_b 
\eean
We first consider the term involving $Q_{ab}^{GR}$. On the horizon, we
have\cite{iyer1} 
\bean
\grad_c\xi_d=\kappa\ep_{cd},
\eean
where $\kappa$ is the surface gravity and $\ep_{cd}$ is the binormal
to $\sh$ (See \cite{iyer1} for further details). Then
\bean
\int_{\sh}Q_{ab}^{GR}[\xi]=\frac{1}{8\pi}\kappa A \label{qgab}
\eean
where $A$ is the area of the black hole. Remember that $\xi^a$ is a fixed
background quantity relative to the variation ``$\dbar$.'' Using the identity
\bean
\dbar\int_{\sh} Q_{ab}^{GR}[\xi]=\delta\int_{\sh}
Q_{ab}^{GR}[\xi]-\int_{\sh} Q_{ab}^{GR}[\delta \xi],  \label{id}
\eean
we have 
\bean
\dbar\int_{\sh} Q_{ab}^{GR}[\xi]&=&\frac{1}{8\pi}\delta (\kappa
A)+\frac{1}{16\pi} \int_{\sh} \ep_{abcd} \grad^c \delta\xi^d
\nonumber \\
&=&
\frac{1}{8\pi}\delta (\kappa
A)+\frac{\delta\Omega_H}{16\pi} \int_{\sh} \ep_{abcd} \grad^c
\varphi^d   \nonumber \\
&=&
\frac{1}{8\pi}\delta (\kappa A)+\delta\Omega_H J_H, \label{dqg}
\eean
where Eqs. \eq{qgab} and \eq{gauge} were used and
$J_H\equiv 1/16\pi \int_{\sh} \ep_{abcd} \grad^c  \varphi^d$
can be interpreted as the angular momentum of the black
hole \cite{gr}. The computation in \cite{bardeen} reveals 
\bean
&&\int_{\sh} \xi\cdot\mf\Theta^{GR}  \nonumber \\\nonumber
&=&\frac{1}{16\pi} \int_{\sh} \xi^a \ep_{dabc}g^{dh}(\grad^e\delta 
g_{he}-g^{fe}\grad_h \delta g_{fe} )   \nonumber \\
&=&\frac{1}{8\pi}A\delta \kappa+\delta\Omega_H J_H. \label{qabi}
\eean
Thus, combining Eqs, \eq{dqg} and \eq{qabi}, we have
\bean
\int_{\sh} \dbar\mf Q^{GR}-\xi\cdot \mf \Theta^{GR}= \frac{1}{8\pi}\kappa
\delta A.  \label{cgr} 
\eean
This result can be viewed as the net contribution from the gravitational field.
We now consider the EM field. By using the smoothness of the pullback
of $A_a$ and the stationary 
condition, one can show that $\Phi^{EM} \equiv -\xi^a
A_a|_{\cal H}$ is a constant in the portion of the horizon to the
future of the bifurcation surface \cite{mine}. If
$A_a$ is smooth over the entire horizon,  $\Phi^{EM}$ will be
identically zero on the horizon since $\xi^a$ vanishes on the
bifurcation surface (in this case, the result in \cite{iyer1} is
recovered).  Together with Eq. \eq{qem}, we have
\bean
\int_{\sh} Q_{ab}^{EM}= \frac{\Phi^{EM} }{8\pi} \int_{\sh} \ep_{abcd} F^{cd}.
\eean
In the asymptotic region, the total electric charge can be expressed
as \cite{gr} 
\bean
\frac{1 }{8\pi} \int_\infty \ep_{abcd} F^{cd}=Q.
\eean
Since the Einstein-Maxwell Lagrangian we considered corresponds to the
sourceless electromagnetic field, the same result must hold if the
integral is performed on the horizon. Therefore
\bean
\int_{\sh} Q_{ab}^{EM}=\Phi^{EM} Q.
\eean
Similar to the identity in Eq. \eq{id}, we have 
\bean
&&\dbar\int_{\sh} Q_{ab}^{EM}[\xi] \nonumber \\
&=&\delta\int_{\sh}
Q_{ab}^{EM}[\xi]-\int_{\sh} Q_{ab}^{EM}[\delta \xi]  \nonumber \\
&=&\delta(\Phi^{EM}Q)+\frac{1}{8\pi}\delta\Omega_H\int_{\sh}\ep_{abcd}F^{cd}A_e
\varphi^e.  
\label{idt}
\eean
Now we compute 
\bean
\int_{\sh}
\xi\cdot\mf\Theta^{EM}=-\frac{1}{4\pi}\int_{\sh}\ep_{cdab}F^{ce}\xi^d \delta
A_e. \label{xth}
\eean
We first express the volume element in the form
\bean
\ep_{cdab}=\xi_c\wedge N_d\wedge \ep_{ab},
\eean
where $\ep_{ab}$ is the volume element on ${\sh}$ and $N^a$ is the
``ingoing'' future directed null normal to ${\sh}$, normalized so
that $N^a\xi_a=-1$ \cite{gr}. Thus, we have
\bean
\int_{\sh}
\xi\cdot\mf\Theta^{EM}=\frac{1}{4\pi}\int_{\sh}\ep_{ab}F^{ce}\xi_c\delta
A_e, \label{xxth}
\eean
By using the fact that on the horizon
$F^{ce}\xi_c\propto\xi^e$\cite{mine}, together with $N^a\xi_a=-1$, we
get immediately 
\bean
F^{ce}\xi_c=F^{cf}N_c \xi_f\xi^e,
\eean
and hence
\bean
\int_{\sh}
\xi\cdot\mf\Theta^{EM}=\frac{1}{4\pi}\int_{\sh}\ep_{ab}F^{cf}N_c
\xi_f\xi^e \delta A_e. \label{xidt}
\eean
On the other hand, 
\bean
Q\delta
\Phi^{EM}&=&-\frac{1}{8\pi}\int_{\sh}\ep_{abcd}F^{cd}\delta(A_e\xi^e)
    \nonumber \\
&=& -\frac{1}{8\pi}\int_{\sh}\ep_{abcd}F^{cd}(\delta
    A_e)\xi^e-\frac{\delta\Omega_H 
    }{8\pi}\int_{\sh}\ep_{abcd}F^{cd}A_e\varphi^e   \nonumber \\
&=& \frac{1}{8\pi}\int_{\sh}  \ep_{ab}\wedge N_c\wedge\xi_d
    F^{cd}\xi^e \delta A_e
    -\frac{\delta\Omega_H}{8\pi}\int_{\sh}\ep_{abcd}
    F^{cd}A_e\varphi^e \nonumber \\
&=& \frac{2}{8\pi}\int_{\sh}  \ep_{ab} F^{cd} N_c\xi_d  \xi^e \delta
    A_e
-\frac{\delta\Omega_H 
    }{8\pi}\int_{\sh}\ep_{abcd}F^{cd}A_e\varphi^e \nonumber \\
&=& \int_{\sh} \xi\cdot\mf\Theta^{EM}-\frac{\delta\Omega_H}
{8\pi}\int_{\sh}\ep_{abcd}F^{cd}A_e\varphi^e 
\label{qdpb}
\eean
Using Eq. \eq{idt}, we have
\bean
\int_{\sh} \dbar Q_{ab}^{EM}-\xi\cdot\mf\Theta^{EM}=\Phi^{EM}\delta Q
\label{qemf} 
\eean
Substitution of Eqs. \eq{qemf} and \eq{cgr} into the right-hand side of
Eq. \eq{lefr} yields Eq. \eq{first}, the desired first law of black hole
mechanics in Einstein-Maxwell theory. As pointed out in the
Introduction section, the potential-charge term \eq{qemf} would
have vanished if the EM field were smooth on the horizon and the integral
were performed on the bifurcation surface. 

\section{The first law in EYM theory}
In this section, we shall extend our derivation in the previous
section to the EYM case. The assumptions and arguments will be similar
to those in the previous section. The EYM Lagrangian takes the form
\bean
\mf L =\frac{1}{16\pi} \mbox{\boldmath$\ep$}  R-\frac{1}{16\pi}
\mbox{\boldmath$\ep$} 
  g^{ac}g^{bd}F_{ab}^\Lambda F_{cd\Lambda}, \label{eyml} 
\eean
where $F_{ab}^\Lambda$ is the Yang-Mills field strength:
\bean
F_{ab}^\Lambda=2\grad_{[a]A_{b}}^\Lambda +c^{\Lambda}_{\ \Gamma\Delta
}A_a^\Gamma A_b^\Delta,  \label{str}
\eean
where $c^{\Lambda}_{\ \Gamma\Delta}$ denotes the structure tensor for
the SU(2) Lie algebra and the Lie algebra indices are raised and
lowered with the Killing metric $g_{\Gamma\Sigma}=-\frac{1}{2}c^\La_{\
  \Gamma\Sigma}c^\Sigma_{\ \Sigma\La}$. 

Similarly to the the EM case, the Lagrangian can be split into ``GR''
and ``YM'' parts. The contribution from the YM field gives 
\bean
\theta^{YM}_{bcd}&=& -\frac{1}{4\pi}\ep_{abcd}F^{ae}_\Delta \delta
A^\Delta_e, \label{thym} \\
Q_{ab}^{YM}&=&-\frac{1}{8\pi} \ep_{abcd}F^{cd}_\La A_e^\La \xi^e. \label{qym}
\eean
Then
\bean
\int_\infty Q[t]=-\frac{1}{8\pi}\int_\infty \ep_{abcd}F^{cd}_\La
A^\La_0. \label{qtym}
\eean
We choose a stationary solution of the EYM equations and then
$A^\La_0$ is asymptotically constant \cite{sw}. The constant $V$ is
defined by
\bean
V=\lim_{r\rightarrow \infty}(A_0^\La A_{0\La})^{1/2} \label{defv}
\eean
The electric field, viewed as a tensor density of weight, is 
\bean
E^a_\La =\sqrt{h} F_{\mu\La}^{\ a} n^\mu, \label{deee}
\eean
where $n^\mu$ is the unit normal to the spacelike hypersurface
$\infty$. Reference \cite{sw} shows that, asymptotically, $A_0^\La$ and
$E_a^\La$ point in the same Lie algebra direction and therefore
\bean
\int_\infty Q[t]=V Q^\infty, \label{qtvi}
\eean
where the Yang-Mills charge measured at infinity is defined by 
\bean
Q^\infty=\frac{1}{4\pi}\int_\infty|E^a_\La r_a|
\eean
where $r^a$ denotes the unit radial vector and vertical bars denote the
Lie algebra norm. On the other hand, 
\bean
&&\int_\infty t\cdot\theta^{YM} \nonumber \\
&=&-\frac{1}{4\pi}\int_\infty \ep_{abcd}t^b F_\Delta^{ae}\delta A_e^\Delta
\nonumber \\
&=&\frac{1}{4\pi}\int_\infty E_\Delta^{a}r_a\delta A_0^\Delta \nonumber \\
&=& Q^\infty \delta V, \label{tdt}
\eean
Therefore, the ``YM'' contribution to $\delta {\cal E}$ is
\bean
\delta {\cal E}_{YM}=V\delta Q^\infty. \label{deym}
\eean
Since the ``GR''contribution gives the ADM mass $M$, we have the total
variation of the canonical energy
\bean
\delta {\cal E}=\delta M+V\delta Q^\infty, \label{tce}
\eean
which agrees with the result in \cite{sw}. 

By using the arguments parallel to that in section 3, we obtain an
expression similar to Eq. \eq{lefr}
\bean
\delta {\cal E} -\Omega_H\delta {\cal J} =\int_{\sh}(\dbar \mf
Q[\xi]-\xi\cdot\mf \Theta), \label{lefy}
\eean
Note that the ADM mass on the left-hand side of Eq. \eq{lefr} has been
replaced by $ {\cal E}$. The canonical angular momentum $J$ is
defined by \cite{iyer1}
\bean
J=-\int_\infty \mf Q[\varphi]. \label{daj}
\eean
Combining Eqs. \eq{qgr} and \eq{qym}, we have
\bean
J=\frac{1}{16\pi}\int_\infty \ep_{abcd}\grad^c\xi^d
+\frac{1}{8\pi} \int_\infty  
\ep_{abcd}F^{cd}_\La A_e^\La \xi^e. \label{cdj}
\eean
This formula agrees with that in \cite{sw}. This first term is just
the expression for angular momentum in the vacuum case.

Since Eq. \eq{cgr} also holds for the EYM case,
we use it to rewrite the right-hand side of Eq. \eq{lefy}
\bean
\delta {\cal E} -\Omega_H\delta J =\frac{1}{8\pi}\kappa
\delta A +\int_{\sh}(\dbar \mf
Q^{YM}[\xi]-\xi\cdot\mf \Theta^{YM}, \label{ale})
\eean
The same treatment used for the EM field gives
\bean
&&\dbar\int_{\sh} Q_{ab}^{YM}[\xi] \nonumber \\
&=&\delta\int_{\sh}
Q_{ab}^{YM}[\xi]-\int_{\sh} Q_{ab}^{YM}[\delta \xi]  \nonumber \\
&=& -\frac{1}{8\pi}\delta \int_{\sh} \ep_{abcd}F^{cd}_\La A_e^\La
\xi^e +\frac{1}{8\pi}\delta\Omega_H\int_{\sh}\ep_{abcd} 
F^{cd}_\La A_e^\La 
\varphi^e \nonumber \\
&=& -\frac{1}{8\pi} \int_{\sh} A_e^\La
\xi^e \delta(\ep_{abcd}F^{cd}_\La )-\frac{1}{8\pi}\int_{\sh}
\ep_{abcd}F^{cd}_\La \delta(A_e^\La \xi^e)
+\frac{1}{8\pi}\delta\Omega_H\int_{\sh}\ep_{abcd}  
F^{cd}_\La A_e^\La 
\varphi^e  \nonumber. \\
\label{idl}
\eean
Replacing the second term of Eq. \eq{idl} by an expression analogous
to Eq. \eq{qdpb}, we get 
\bean
\dbar\int_{\sh} Q_{ab}^{YM}[\xi]=-\frac{1}{8\pi} \int_{\sh} A_e^\La
\xi^e \delta(\ep_{abcd}F^{cd}_\La )+\int_{\sh}\xi\cdot{\mf
  \Theta}^{YM}. \label{oid}
\eean
Then, from Eqs. \eq{ale} and \eq{oid}, we obtain the first law for a
stationary EYM black hole:
\bean
\frac{1}{8\pi}\kappa \delta A=\delta {\cal E} -\Omega_H\delta
J-\frac{1}{8\pi} \int_{\sh} A_e^\La \xi^e 
\delta(\ep_{abcd}F^{cd}_\La ). \label{firym}
\eean
This expression agrees with that in \cite{sta}. We cannot further
evaluate the integral in the form of ``$\Phi\delta Q$''  as in the EM case
because of the complicity of SU(2) Lie algebra. Ashtekar, {\em
  et. al.}\cite{ash}   
chose the following gauge conditions (see also Corichi, {\em et. al.}
\cite{acor}). 
 
(i) The Yang-Mills potential 
\bean
\Phi^{YM}=-|\xi\cdot \mf A| \label{dyp}
\eean
is constant on the horizon. 

(ii) The dual of the field strength $(\mf{^*F})$ and $(\xi\cdot \mf A)$
  point in the same Lie algebra direction
\bean
(\xi\cdot \mf A)^\Sigma\propto (^2\ep\cdot \mf{^*F})^\Sigma, \label{cot}
\eean
where $^2\ep$ is the pullback to the horizon of $\ep_{abcd}$. Under
  these two conditions, the integral in Eq. \eq{firym} can be evaluated as
\bean
-\frac{1}{8\pi} \int_{\sh} A_e^\La \xi^e
\delta(\ep_{abcd}F^{cd}_\La )= \Phi^{YM}\delta Q^{YM}_{{\cal H}}  \label{far}
\eean
where $ Q^{YM}_{{\cal H}}=-(1/4\pi)\int_{\sh} |\mf{^*F}|$ is
the electric Yang-Mills charge evaluated on the horizon. However,
  there is no evidence that our 
  stationary gauge choice is consistent with conditions (i) and (ii)
  above. Therefore, Eq. \eq{firym} is our final form of the first law in EYM
  theory.  

\section{Conclusions}
The first law of black hole mechanics for the EM and EYM cases is
derived in the framework of \cite{iyer1}. In contrast to \cite{iyer1},
we make no reference to the bifurcation surface. In the EM case, when
the pullback of $A_a$ to the future horizon is smooth, the desired
charge-potential term is obtained. In the EYM case, a 
corresponding surface integral on the horizon is found. Since we avoid
using the bifurcation surface, the derivation and conclusions in this
paper apply to
extremal black holes simply by taking $\kappa=0$.   

\noindent
\begin{center}
{\bf  \large Acknowledgments} 
\end{center}

I would like to thank professor Robert M. Wald for helpful comments on the
manuscript. I also wish to thank Dr. Xiaoning Wu for useful
discussions. This work was supported in part by FCT (Portugal).

\end{document}